# Quest for New Quantum States via Field-Editing Technology

Gang Cao[1*], Hengdi Zhao[1], Bing Hu[1,2], Nicholas Pellatz[1], Dmitry Reznik[1], Pedro Schlottmann[3] and Itamar Kimchi[1]

[1]Department of Physics, University of Colorado at Boulder, Boulder, Colorado 80309, USA

[2]School of Mathematics and Physics, North China Electric Power University, Beijing 102206, China

[3]Department of Physics, Florida State University, Tallahassee, Florida 32306, USA

We report new quantum states in spin-orbit-coupled single crystals that are synthesized using a game-changing technology that *"field-edits"* crystal structures (*borrowing from the phrase "genome editing"*) via application of magnetic field during crystal growth. This study is intended to fundamentally address a major challenge facing the research community today: A great deal of theoretical work predicting exotic states for strongly spin-orbit-coupled, correlated materials has thus far met very limited experimental confirmation. These conspicuous discrepancies are due chiefly to the extreme sensitivity of these materials to structural distortions. The results presented here demonstrate that the "*field-edited*" materials not only are much less distorted but also exhibit novel phenomena absent in their "*non-edited*" counterparts. The field-edited materials include an array of *4d* and *5d* transition metal oxides, and three representative materials presented here are $Ba_4Ir_3O_{10}$, $Ca_2RuO_4$, and $Sr_2IrO_4$. This study provides an entirely new approach for discovery of new quantum states and materials otherwise unavailable.

*Corresponding author: gang.cao@colorado.edu

The discovery of novel quantum states requires innovative approaches. Here, we report new quantum states in single crystals of *4d-* and *5d*-electron based oxides obtained through structurally "editing" materials *(borrowing from the phrase, "genome editing")* via application of magnetic field during materials growth -- A magnetic field aligns magnetic moments and, through strong spin-orbit interactions (SOI) and magnetoelastic coupling, "edits" crystal structures at high temperatures, as schematically illustrated in **Fig.1a**. This work demonstrates that such a ***"field-editing" technology*** is highly effective for quantum materials with a *delicate interplay* between SOI and comparable Coulomb correlations, in which new quantum states readily arise whenever competing interactions conspire to provoke unusually large susceptibilities to small, external stimuli [1-7].

The field-editing technology we have developed directly addresses a major challenge to today's research community: A great deal of theoretical work that predicts exotic states for correlated and spin-orbit-coupled oxides has thus far met very limited experimental confirmation [1, 2, 6-15]. We have long recognized that the conspicuous discrepancies between current theory and experiment are due chiefly to the extreme susceptibility of relevant materials to structural distortions and disorder, a key characteristic existent in strongly spin-orbit coupled matter [2,3]. The field-editing technology fundamentally addresses this challenge by "field-editing" materials via application of magnetic field during materials growth at high temperatures (**Fig.1a**), and this study demonstrates that it works particularly well for materials that are dictated by competing and comparable fundamental interactions [1-3, 6-15] and inherently have multiple *nearly degenerate states* (loosely analogous to a conventional "spin glass" existent in a triangular spin lattice). Consequently, even small perturbations or slightly "editing" distortions/disorder can provoke unusually strong responses in physical properties or remove the "degeneracy", facilitating a new ground state to emerge [2,6,16-24]. In particular, the applied magnetic field during crystal growth exerts a torque on the magnetic moments, which via strong SOI can change the bond angles, the overlap matrix elements of the orbitals and hence the physical properties. It is astonishing that all these drastic changes in physical properties are a consequence of field-editing with a very weak magnetic field no stronger than 0.06 Tesla, as exampled in **Figs.1b-1c** (discussed below). This is utterly inconsistent with conventional thermodynamics, according to which even an extremely strong magnetic field (e.g., 45 Tesla ~ 4 meV) would seem inconsequential to chemical reactions as magnetic contributions to the Gibbs free enthalpy are too small to be significant in terms of energy scale [25]. Indeed, effects of growing silicon in weak magnetic fields were limited [26]. This work puts forward an entirely new approach for an otherwise unavailable path to discovery of novel quantum states/materials based on competing interactions.

Here, we report results of our study on three representative single-crystal materials that are "field-edited" and our controlled study on their "non-edited" counterparts. These three representative materials are the *5d*-electron based iridates $Ba_4Ir_3O_{10}$ [27], $Sr_2IrO_4$ [28] and the *4d*-electron based ruthenate $Ca_2RuO_4$ [29]. This comprehensive investigation involves an array of *4d* and *5d* oxides, and results of other studied materials will be presented in a separate paper. The study of all these materials reveals a common empirical trend that the field-edited single-crystal materials are much less distorted than their non-edited counterparts, and exhibit quantum states that are either absent or vastly different from those in the non-edited materials (e.g. **Figs.1b-1c**).

Experimental details including the single crystal synthesis, single-crystal x-ray diffraction, energy dispersion x-ray and measurements of physical properties are presented in the Supplemental Material [30]. A few crucial details are worth mentioning here. The field-edited single crystals are grown in a 1500 °C-furnace carefully surrounded with two specially-made permanent magnets, each of which is of 1.4 Tesla (**Fig.1a**). Since the magnetic field of a permanent magnet decays with distance $d$ as $1/d^3$, the actual strength of the magnetic field inside the furnace chamber is measured to be within a range of 0.02 - 0.06 Tesla. The non-edited single crystals are synthesized without the applied magnetic field in otherwise identical conditions. All results reported here are repeatedly confirmed by samples from multiple batches of single crystals synthesized throughout the nearly one-year period of this study.

It is emphasized that the work presented here can serve as proof-of-concept results; the field-editing technology with much stronger magnetic fields and higher temperatures will undoubtedly result in more discoveries of novel quantum states and materials.

For contrast and comparison, the structural and physical properties of both field-edited and non-edited samples are simultaneously presented.

### A. $Ba_4Ir_3O_{10}$: From Quantum Liquid to Correlated Antiferromagnet

The magnetic insulator $Ba_4Ir_3O_{10}$ is recently found to be a novel quantum liquid [31]. As shown in **Fig.2**, $Ba_4Ir_3O_{10}$, which adopts a monoclinic structure with a $P2_1/c$ space group, is structurally a two-dimensional, square lattice with no apparent spin chains. Our recent study reveals the quantum liquid persisting down to 0.2 K that is stabilized by strong antiferromagnetic (AFM) interaction with Curie-Weiss temperature ranging from -766 K to -169 K due to magnetic anisotropy. The anisotropy-averaged frustration parameter, defined as $f= |\theta_{CW}|/T_N$, is more than 2000, seldom seen in other materials. Heat capacity and thermal conductivity are both linear at low temperatures, a defining characteristic for an exotic quantum liquid state. The novelty of the state is that frustration occurs in an un-frustrated square lattice which features $Ir_3O_{12}$ trimers of face-sharing $IrO_6$ octahedra. It is these trimers that form the basic magnetic unit and play a crucial role in frustration. In particular, a combined effect of the direct (Ir-Ir) and superexchange (Ir-O-Ir) interactions in the trimers results in such a delicate coupling that the middle Ir ion in a trimer is only very weakly linked to the two neighboring Ir ions. Such "weak-links" generate an effective one-dimensional system with zigzag chains or Luttinger liquids along the *c* axis [31].

This intricacy is fundamentally changed in the field-edited $Ba_4Ir_3O_{10}$. Structurally, the field-edited single crystal exhibits a significant elongation in the *b* axis with only slight changes in the *a* and *c* axis, compared to those of the non-edited sample. As a result, the unit cell volume V increases considerably by up to 0.54% at 350 K (see **Figs 2a-2c**). Remarkably, both the Ir-Ir bond distance within each trimer and the Ir-O-Ir bond angle between trimers increase sizably, as schematically illustrated in **Figs.2f-2g**.

The effect of field-editing not only readily destroys the quantum liquid inherent in the non-edited samples but also stabilizes a robust, long-range magnetic order. As shown in **Fig.3a**, two magnetic anomalies occur at Neel temperatures $T_N$=125 K and $T_{N2}$=12 K in the field-edited sample (solid dots), sharply contrasting the magnetic behavior of the non-edited sample (dashed lines). Consequently, the absolute values of the Curie-Weiss temperature $\theta_{CW}$ are considerably reduced and become comparable to $T_N$=125 K for the field-edited sample (**Figs.3b-3c**); the corresponding frustration parameter f ($=|\theta_{CW}|/T_N$) is accordingly reduced to a value less than 3, a drastic drop from the average value of 2000 in the non-edited sample [31], indicating a complete removal of frustration. The long-range magnetic order is corroborated by a metamagnetic transition at a critical field $H_c$ = 2.5 T along the *a* axis observed in the isothermal magnetization M(H) (**Fig.3d**). Metamagnetism often occurs in canted AFM states in quasi-two-dimensional lattices [32].

The heat capacity, which measures bulk effects, confirms the AFM order. In particular, the low-temperature linearity of the heat capacity C(T) (data in blue **Fig.3e**), which characterizes the gapless excitations in the non-edited sample, is replaced by the $T^3$-dependence in the field-edited sample, which is anticipated for an insulating antiferromagnet (data in red in **Fig. 3e**). Along with the linearity of C(T), the sharp upturn in C(T) at T* = 0.2 K in the non-edited sample also disappears in the field-edited sample. These changes clearly illustrate that the ground state of $Ba_4Ir_3O_{10}$ is fundamentally changed!

Indeed, as temperature rises, two anomalies occur at $T_{N2}$ = 12 K (**Fig.3f**) and $T_N$ = 125 K (**Fig.3g**), respectively, confirming the robustness of the long-range magnetic order observed in the magnetic data in **Fig.3a-3d**.

The observed magnetic order in the field-edited sample is similar to that observed in slightly doped $Ba_4Ir_3O_{10}$, in which a mere 2% Sr substitution for Ba produces long-range order at 130 K [31]. However, remarkable differences are in the details. Most notably, the anomaly at $T_{N2}$ (= 12 K) in C(T) of the field-edited samples is absent in C(T) of the Sr-doped sample, and the metamagnetic transition for $M_a$ occurs at $H_c$ = 2.5 T and 4.2 T for the field-edited and Sr-doped sample, respectively. Furthermore, $M_b$ for the Sr-doped sample shows an additional $H_c$ at 6.5 T, which is absent in $M_b$ for the field-edited sample.

In short, the quantum liquid in the non-edited $Ba_4Ir_3O_{10}$, which is attributed to the reduced intra-trimer exchange and weakly coupled one-dimensional chains along the *c* axis [31], is replaced in the field-edited $Ba_4Ir_3O_{10}$ by the strongly AFM state stabilized by three-dimensional correlations.

### B. $Ca_2RuO_4$: From Collinear Antiferromagnet to Weak Ferromagnet

The AFM insulator $Ca_2RuO_4$ exhibits a metal-insulator transition at $T_{MI}$ = 357 K [29,33], which marks a concomitant, violent structural transition with a severe rotation and tilt of $RuO_6$. This structural transition removes the $t_{2g}$ orbital degeneracy ($d_{xy}$, $d_{yz}$, $d_{zx}$), which dictates the physical properties of $Ca_2RuO_4$ [22, 34 - 44]. An AFM transition occurs only at a considerably lower Neel temperature $T_N$=110 K [29], highlighting its close association with a further distorted structure. Extensive investigations of this system have established that quantum effects are intimately coupled to lattice perturbations [34 - 44].

As shown in **Fig. 4**, the crystal structure of $Ca_2RuO_4$ is significantly field-edited, becoming less distorted. A few changes are particularly remarkable. The first-order structural transition $T_{MI}$ is suppressed by about 25 K from 357 K to 332 K, which is marked by the vertical blue dashed and red solid lines, respectively, through **Figs. 4a-4c.** In the field-edited structure, the *c* axis gets longer (**Fig. 4a**); the *b* axis becomes shorter whereas the *a* axis changes very slightly (**Fig. 4b**), thus leading to a reduced orthorhombicity (**Fig. 4d**). Furthermore, the O2-Ru1-O2 and Ru1-O2-Ru1 bond angles, which measure the octahedral rotation and tilt, get relaxed, in the field-edited structure **(Fig. 4e-4f)**. All these lattice changes are critical to both transport and magnetic properties. The crystal structure in the *ac* and *ab* planes and the schematic for the bond angles are shown in **Fig.4g-4i.**

Indeed, the *a*-axis electrical resistivity $\rho_a$ of the field-edited sample shows a much lower metal-insulator transition $T_{MI}$ at 324 K, 31 K lower than $T_{MI}$ for the non-edited sample, as seen in **Fig. 5a**. The suppressed $T_{MI}$ closely tracks the structural transition that is reduced by about 25 K in the field-edited sample (**Fig.4**).

Magnetically, the field-edited sample behaves vastly differently from the non-edited sample. In particular, the *a*-axis magnetic susceptibility $\chi_a$ of the field-edited sample shows a ferromagnetic-like behavior with the onset of the magnetic transition at $T_N$ = 135 K (red curves), in sharp contrast to that of the non-edited sample (blue curve) (see **Fig. 5b**). The increased $T_N$ is likely related to a sizable increase in the unit cell V below 200 K via the strong magnetoelastic coupling (see **Inset** in **Fig. 4c**). Moreover, a large hysteresis behavior of $\chi_a$ is observed in the field-edited sample (**Inset** in **Fig. 5b**), which is absent in the non-edited sample but expected in a ferromagnet or weak ferromagnet. In this case, the field-edited sample more likely becomes a weak ferromagnet or canted antiferromagnet; this is consistent with a metamagnetic transition, $H_c = 2.4$ T, observed in the isothermal magnetization M(H) illustrated in **Fig. 5c**. The non-edited $Ca_2RuO_4$ is a known collinear antiferromagnet without any metamagnetic behavior [3, 29]. The magnetic changes are also in accordance with changes in the low-temperature heat capacity C(T). For an insulating antiferromagnet, $C(T) \sim (\alpha + \beta) T^3$, in which the first term $\alpha$ and the second term $\beta$ are associated with magnon and phonon contributions to C(T), respectively. Here, C(T) shows a significant slope change defined by $(\alpha + \beta)$ in the plot of $C/T$ vs $T^2$ in **Fig. 5d**. Such a slope change clearly points out that the emergent magnetic state is distinctly different from the native AFM state, consistent with the magnetic data in **Figs.5b-5c**.

### C. $Sr_2IrO_4$: Towards Novel Superconductivity

$Sr_2IrO_4$ is an archetype of the spin-orbit-driven magnetic insulator [45,46], an extensively studied material in recent years [2, 3, 6-12]. It is widely anticipated that with slight electron doping, $Sr_2IrO_4$ should be a novel superconductor [2, 9-12, 24]. However, there has been no experimental confirmation of superconductivity, despite many years of experimental effort [2]. We believe that the absence of the predicted superconductivity is due to *inherently severe structural distortions* that suppress superconductivity [2, 16-18, 47]. This point is also supported by a recent theoretical study, which attributes the lack of superconductivity to the octahedral rotation [24]. In fact, it is precisely because of this early realization that we initiated the development of the field-editing technology and investigations of field-edited materials. Indeed, the structural, magnetic and transport properties of the field-edited $Sr_2IrO_4$ and 3% La doped $Sr_2IrO_4$ or $(Sr_{0.97}La_{0.03})_2IrO_4$ are either drastically improved or changed, compared to those of the non-edited samples. In particular, the field-edited structure is more expanded and less distorted (Ir-O-Ir bond angle becomes larger) (**Figs.6a-6b**), and the AFM transition $T_N$ is suppressed by astonishing 90 K (**Fig.6c**); the isothermal magnetization is reduced by 50% and much less "saturated" compared to that for the non-edited $Sr_2IrO_4$ (see Supplementary Fig.1[30]). That the Ir-O-Ir bond angle dictates $T_N$ and magnetization suggests a critical role of the Dzyaloshinskii-Moriya interaction, consistent with early study of the iridate [6]. Indeed, such magnetic changes are clearly reflected in Raman scattering. One-magnon Raman scattering measures the anisotropy field that pins the magnetic moment orientation. It broadens with increasing temperature and vanishes at $T_N$. At 10 K, this peak in the non-edited $Sr_2IrO_4$ occurs near 18 cm$^{-1}$ (data in blue in **Fig.6d**) [48] but is absent in the field-edited $Sr_2IrO_4$ for the measured energy range (data in red **Fig.6d**). This conspicuous disappearance of the peak clearly indicates that the anisotropy field is drastically reduced and, consequently, the one-magnon peak is either completely removed or suppressed to an energy below the energy cutoff of 5.3 cm$^{-1}$ (0.67 meV) in the field-edited sample. On the other hand, two-magnon scattering remains essentially unchanged (see Supplementary Fig.2 [30]).

Furthermore, *the resistivity is reduced by up to five orders of magnitude and shows a nearly metallic behavior at high temperatures in the field-edited $Sr_2IrO_4$* (**Figs.6e-6f**)! Also note that there is an anomaly corresponding to $T_N$ = 150 K (see **Inset** in **Fig.6e**), indicating a close correlation between the transport and magnetic properties that is noticeably absent in the non-edited $Sr_2IrO_4$ (**Figs 6c, 6e-6f**) [2]. This is consistent with the fact that the drastically improved conductivity (**Figs.6e-6f**) is accompanied by the equally drastically weakened magnetic state (**Figs.6c-6d**). It is therefore not surprising that the resistivity for the field-edited $(Sr_{0.97}La_{0.03})_2IrO_4$ exhibits an abrupt drop below 20 K by nearly three orders of magnitude, suggesting that the long-elusive superconductivity in the iridate may be finally within reach (these results are to be reported in a separate paper).

### D. Proposed Proof-of-Concept Theoretical Mechanism

The tiny fields of 200 gauss sufficient for field-editing involve a magnetic Zeeman energy scale of only 0.002 meV per spin-1/2, orders of magnitude smaller than typical electronic energy scales. It is a priori not clear how any mechanism could possibly allow field editing with such small energy scales. Here we explore possible proof of concept mechanisms that would enable such an effect.

Consider the effects of magnetic fields on crystal growth observed in other systems. Crystallization of proteins has been shown to benefit from applied large magnetic fields on the order of a few tesla or more. The mechanisms for this effect are thought to rely on increased viscosity and more uniform alignment within the high temperature fluid [49]. However, the energy landscape for protein crystallization is dominated by the ultraweak van der Waals interactions, such that few-tesla magnetic fields can make a direct impact on the preferred alignments and associated crystallization energies; in contrast, the presently studied transition metal oxide solids are crystallized via covalent and ionic bonding energies which are many orders of magnitude higher. A first distinction between the present study and previous work is thus in the ratio of magnetic field strengths to crystallization energies.

A second distinction is in the effect of field growth. In the literature on semiconductor crystal growth as well as in the protein crystallization cases, it has been shown that magnetic fields enable higher quality crystals or larger crystals [26, 50]. These effects arise from the interface between the high temperature melt and the growing crystal, hence are readily understood in terms of the magnetic field effects in the molten fluid properties. Here in contrast we find that the fields preserve the formation of large bulk crystals but modify the resulting crystal structure, including its electronic properties.

These distinctions suggest that, while the high temperature magnetohydrodynamic effects observed elsewhere are expected to also play a role here, an additional qualitatively new mechanism is necessary to explain the present observations. The question we address here concerns the modifications of the low energy state: what energetic mechanism could allow the weak magnetic field to change the crystal structure so drastically as to change local bond angles by even a few percent, with the associated enormous changes in the electronic and magnetic properties discussed above? We proceed to construct a proof of concept of such a qualitatively new mechanism for field editing, based on the combination of (1) strong spin-orbit or magnetoelastic couplings, and (2) magnetic frustration.

Consider the low energy theory of the magnetic insulators. It can be expressed in terms of effective spin-orbit-coupled spin-1/2 degrees of freedom, which we represent by the Pauli matrices $\tau^\alpha$. The interaction among these degrees of freedom can be written schematically as

$$H = J_{r,r'}^{\alpha\alpha'} \sum_{r,r'} \tau_r^\alpha \tau_{r'}^{\alpha'}$$

The coupling of magnetic field to the crystal structure appears in two ways: in the spin-orbital wavefunction content of $\tau^\alpha$, and in the magnetoelastic dependence on the interaction $J$ on the spatial positions *r,r'*.

Now observe that such Hamiltonians can exhibit large magnetic frustration and associated near-degeneracies among exponentially many quantum states. This can occur even when the underlying magnetic lattice has no geometric frustration; rather the frustration can arise purely from spin-orbit coupling. As a proof-of-concept example, consider the square lattice quantum compass model [51],

$$H_c = J_r^h \sum_{h-\text{bond}} \tau_r^x \tau_{r+x_0}^x + J_r^v \sum_{v-\text{bond}} \tau_r^y \tau_{r+y_0}^y$$

where $h$-bonds and $v$-bonds refer to horizontal and vertical bonds respectively, with the vector along the bond being $x_0$ and $y_0$ respectively. This term has been argued to arise in the effective description of $Sr_2IrO_4$ [6] and is also expected to arise by symmetry considerations for $Ca_2RuO_4$.

The pure Hamiltonian $H_c$ possesses fine-tuned symmetries that give it unusual properties, which here also translate to unusual changes in the crystal and its electronic structure [51]. First, every quantum state of this model, including the ground state, is degenerate with an enormous number of other states, giving a sequence of degenerate manifolds each of which is exponentially large in system size ($2^L$). Second, 1D domain wall defects in the 2D spin system have only a fixed energy cost independent of system size $L$ rather than the usual linear scaling $L$; however their entropy still diverges with system size, giving a free energy $E - TS$ that is infinitely negative relative to the ground state, hence a finite density of domain wall defects at any nonzero temperature. The spin-orbit and magnetoelastic couplings imply that each defect in the effective-spin texture also produces a change in the local orbitals and local distortions of the crystal.

Now consider the effect of applying a magnetic field. The magnetoelastic coupling requires us to go beyond the spin manifold and include ionic positions. Similarly the magnetic field will also couple differently to the orbital and spin characters of the effective spins and thus modify the orbitals, again coupling to the crystal structure. Within a model such as $H_c$, the degeneracy of the manifold containing each state implies that even infinitesimally small magnetic fields have a singular and large effect on the free energy landscape. The manifolds with the lowest free energy (and large defect density relative to the ground state) will be split and rearranged so as to favor states with uniform and nonzero magnetization, with those states forming a complicated landscape of free energy barriers reminiscent of spin glasses. If taken literally, this argument would also suggest that the dynamics of the crystal formation can proceed more easily through states with many proliferated defects, and thus are less likely to lead to any state with a particular long range pattern of distortions; this would be consistent with the apparent experimental result that $Ca_2RuO_4$ and $Sr_2IrO_4$ have less crystalline distortions (such as octahedra tilts) upon field editing, but we caution that this part of the theoretical argument may not generalize beyond the fine tuned $H_c$ Hamiltonian.

The full Hamiltonian has many strong competing terms beyond any single fine-tuned-frustration term such as $H_c$. When the full Hamiltonian still results in magnetic frustration, as indeed is observed in all three unedited compounds discussed above, the frustrated system no longer has degenerate manifolds but instead shows a glass-like landscape of states, some of which lie nearby in energy but have drastically different spin configurations. As the crystal forms and samples this free energy landscape, even small magnetic field magnitudes can have a large effect on the dynamics of the crystal distortion relaxation and electronic structure.

The change in orbital distortions is observable not only in the structural and magnetic properties but also in the electronic hopping and resistivity. The direct observation of this connection was recently argued for $Ca_2RuO_4$ [22] and it is reasonable to conjecture that $Sr_2IrO_4$ will exhibit an analogous but stronger effect due to its stronger spin orbit coupling. Note also that the $H_c$ toy model can be applicable to $Ca_2RuO_4$ and $Sr_2IrO_4$ but not $Ba_4Ir_3O_{10}$; however, the immense frustration seen in unedited $Ba_4Ir_3O_{10}$ suggests that an analogous mechanism is likely at play there, whose understanding will require further microscopic analysis [31].

In conclusion, all results presented above have clearly demonstrated that the field-editing technology is extraordinarily effective for generating new quantum states in correlated and spin-orbit-coupled materials. It is particularly astonishing that all this is achieved via an applied magnetic field no stronger than 0.06 Tesla during materials growth. This is a game-changing technology, and with stronger magnetic fields, it will overcome more materials challenges, leading to more discoveries of novel quantum states and materials that cannot be produced otherwise.

**Acknowledgements** This work is supported by NSF via grants DMR 1712101 and DMR 1903888. Raman scattering work (N.P. and D.R.) was supported by the NSF under Grant No. DMR-1709946. GC is thankful to Drs. Dan Dessau, Minhyea Lee, Feng Ye and Lance DeLong for useful discussions.

13. *"Doping a spin-orbit Mott insulator: Topological superconductivity from the Kitaev-Heisenberg model and possible application to $(Na_2/Li_2)IrO_3$"*, Y. Z. You, I. Kimchi and A. Vishwanath, Phys. Rev. B **86**, 085145 (2012)

14. *"Kitaev-Heisenberg Model on a Honeycomb Lattice: Possible Exotic Phases in Iridium Oxides $A_2IrO_3$"*, J. Chaloupka, G. Jackeli and G. Khaliullin, Phys. Rev. Lett. **105**, 027204 (2010)

15. *"Spin-Orbit Physics Giving Rise to Novel Phases in Correlated Systems: Iridates and Related Materials"*, J. G. Rau, E. K. H. Lee and H. Y. Kee, Annu. Rev. Condens. Matter Phys. **7**, 195 (2016)

16. *"Giant magnetoelectric effect in the $J_{eff}=1/2$ Mott insulator $Sr_2IrO_4$"*, S. Chikara, O. Korneta, W. P. Crummett, L. E. DeLong, P. Schlottmann and G. Cao, Phys. Rev. B **80**, 140407(R) (2009)

17. *"Electron-doped $Sr_2IrO_{4-\delta}$ ($0 \leq \delta \leq 0.04$): Evolution of a disordered $J_{eff}=1/2$ Mott insulator into an exotic metallic state"*, O. B. Korneta, T. F. Qi, S. Chikara, S. Parkin, L. E. De Long, P. Schlottmann and G. Cao, Phys. Rev. B **82**, 115117 (2010)

18. *"Lattice-driven magnetoresistivity and metal-insulator transition in single-layered iridates"*, M. Ge, T. F. Qi, O. B. Korneta, D. E. De Long, P. Schlottmann, W. P. Crummett and G. Cao, Phys. Rev. B **84**, 100402(R) (2011)

19. *"Lattice-Tuned Magnetism of $Ru^{4+}(4d^4)$ Ions in Single-Crystals of the Layered Honeycomb Ruthenates: $Li_2RuO_3$ and $Na_2RuO_3$"*, J. C. Wang, J. Terzic, T. F. Qi, Feng Ye, S. J. Yuan, S. Aswartham, S. V. Streltsov, D. I. Khomskii, R. K. Kaul and G. Cao, *Phys. Rev. B Rapid Comm.* **90** 161110 (R) (2014)

20. *"Electrical Control of Structural and Physical Properties via Spin-Orbit Interactions in $Sr_2IrO_4$"*, G. Cao, J. Terzic, H. D. Zhao, H. Zheng, L. E DeLong and Peter Riseborough, *Phys. Rev. Lett* **120**, 017201 (2018)

21. *"Observation of a pressure-induced transition from interlayer ferromagnetism to intralayer antiferromagnetism in $Sr_4Ru_3O_{10}$"*, H. Zheng, W.H. Song, J. Terzic, H. D. Zhao, Y. Zhang, Y. F. Ni, L. E. DeLong, P. Schlottmann and G. Cao, *Phys. Rev. B* **98**, 064418 (2018)

22. "*Nonequilibrium Orbital Transitions via Applied Electrical Current in Calcium Ruthenates*", Hengdi Zhao, Bing Hu, Feng Ye, Christina Hoffmann, Itamar Kimchi and16

**Captions**

**Fig. 1. Field-Editing Technology: (a)** A schematic for field-editing a crystal structure (left) during crystal growth in the molten-zone sandwiched between the two magnets (right). Contrasting physical properties of two exemplary materials to highlight field-editing effects: **(b)** $Sr_2IrO_4$: the *c*-axis electrical resistivity $\rho_c$ is seven orders of magnitude smaller in the field-edited crystal (red) than in the non-edited counterpart (blue); **(c)** $Ba_4Ir_3O_{10}$: a comparison of the *a*-axis magnetic susceptibility $\chi_a$ between the non-edited crystal (blue) that features a highly frustrated quantum liquid and the field-edited crystal (red) that becomes a robust antiferromagnet with two magnetic transitions.

**Fig. 2. $Ba_4Ir_3O_{10}$: Structural properties of the field-edited (in red) and non-edited (in blue) single crystals:** The temperature dependence of the lattice parameters **(a)** the *a* and *c* axis, **(b)** the *b* axis and **(c)** the unit cell volume V. **(d)** and **(e)** the crystal structure in the *ab* and *bc* plane, respectively. **(f)** The Ir-Ir bond distance within a trimer. **(g)** The Ir-O-Ir bond angle between corner-sharing trimers (the marked values are for 100 K).

**Fig. 3. $Ba_4Ir_3O_{10}$: Physical properties of the field-edited and non-edited single crystals:** The temperature dependence for the *a*, *b* and *c* axis of **(a)** the magnetic susceptibility $\chi(T)$ for the field-edited (solid dots) (Inset: zoomed-in $\chi$ near $T_N$), and the non-edited (dashed lines) samples, and $\Delta\chi^{-1}$ **(b)** for the non-edited single crystal and **(c)** for the field-edited single crystal. **(d)** The isothermal magnetization M(H) at 1.8 K for the field-edited (solid lines) and the non-edited (dashed lines) samples. The temperature dependence of the heat capacity C(T) for the field-edited (red) and the non-edited (blue) samples at **(e)** the lowest temperatures, **(f)** the intermediate temperatures and **(g)** the high temperatures. The inset in (g): the zoomed-in C(T) near $T_N$.

**Fig. 4. $Ca_2RuO_4$: Structural properties of the field-edited (in red) and non-edited (in blue) single crystals:** The temperature dependence of the lattice parameters **(a)** the *c* axis, **(b)** the *a* and *b* axis, **(c)** the unit cell volume V, **(d)** the basal plane orthorhombicity, **(e)** the O2-Ru1-O2 bond angle and **(f)** the Ru1-O2-Ru1 bond angle. **(g)** and **(h)** The crystal structure in the *ac* and *ab* plane,

respectively. **(i)** The schematic for the O2-Ru1-O2 and Ru1-O2-Ru1 bond angles for the field-edited (red) and non-edited (blue) structures at 250 K.

**Fig. 5. $Ca_2RuO_4$: Physical properties of the field-edited (in red) and non-edited (in blue) single crystals:** The temperature dependence of **(a)** the *a*-axis electrical resistivity $\rho_a$ and **(b)** the *a*-axis magnetic susceptibility $\chi_a(T)$ at $\mu_oH$ =0.1 T. Inset: $\chi_a(T)$ for the field-cooled (FC) and zero-field cooled (ZFC) sequences for the field-edited sample. **(c)** The *a*-axis isothermal magnetization $M_a(H)$ at 30 K. **(d)** The low-temperature heat capacity C(T) plotted as C/T vs $T^2$.

**Fig. 6. $Sr_2IrO_4$: Structural and physical properties of the field-edited (in red) and non-edited (in blue) single crystals:** The temperature dependence of **(a)** the unit cell volume V, **(b)** the Ir-O-Ir bond angle (Inset illustrates the Ir-O-Ir bond angle θ) and **(c)** the *a*-axis magnetization $M_a$ at $\mu_oH$ =0.1 T. **(d)** The Raman single-magnon peak at 10 K. The temperature dependence of **(e)** the *a*-axis resistivity $\rho_a$ and **(f)** the *c*-axis resistivity $\rho_c$ for $Sr_2IrO_4$. Inset illustrates the anomaly in $\rho_a$ at $T_N$ = 150 K for the field-edited crystal.

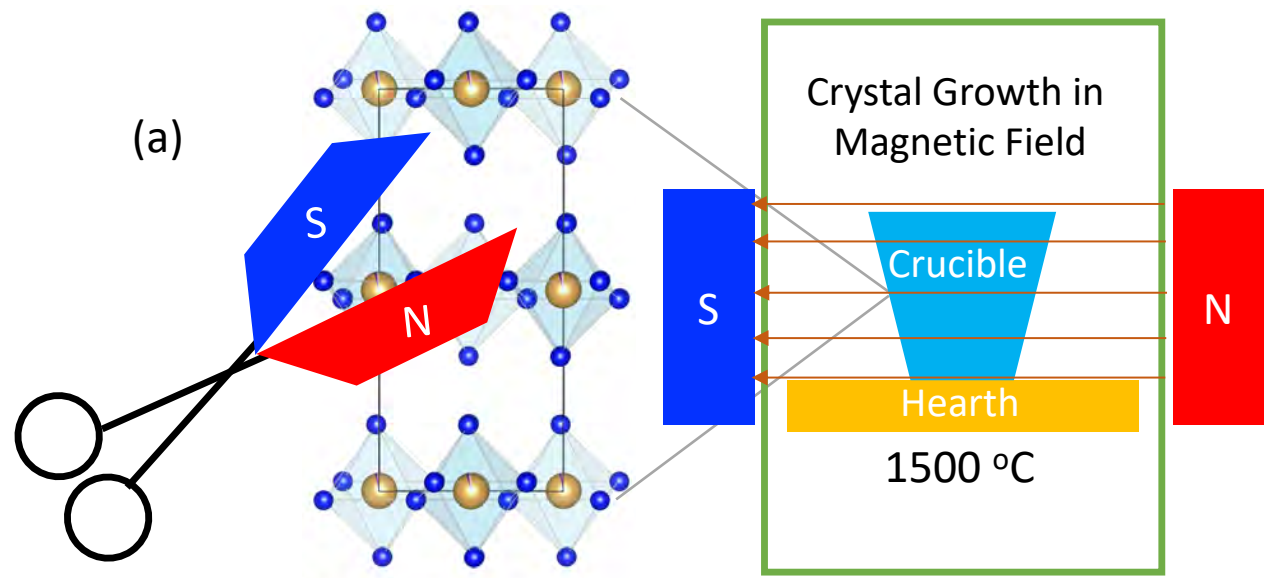
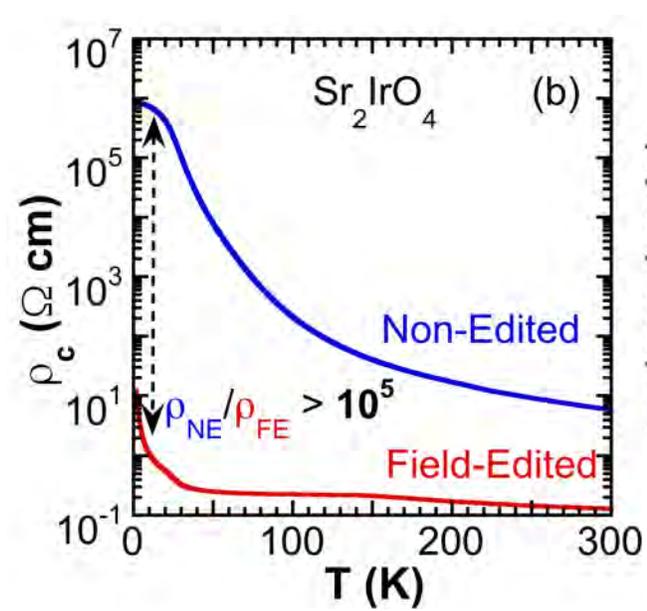
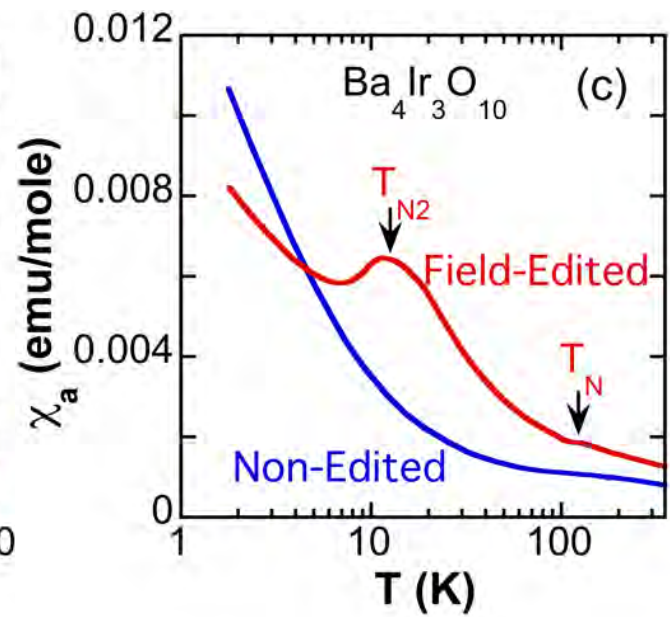

**Fig.1**

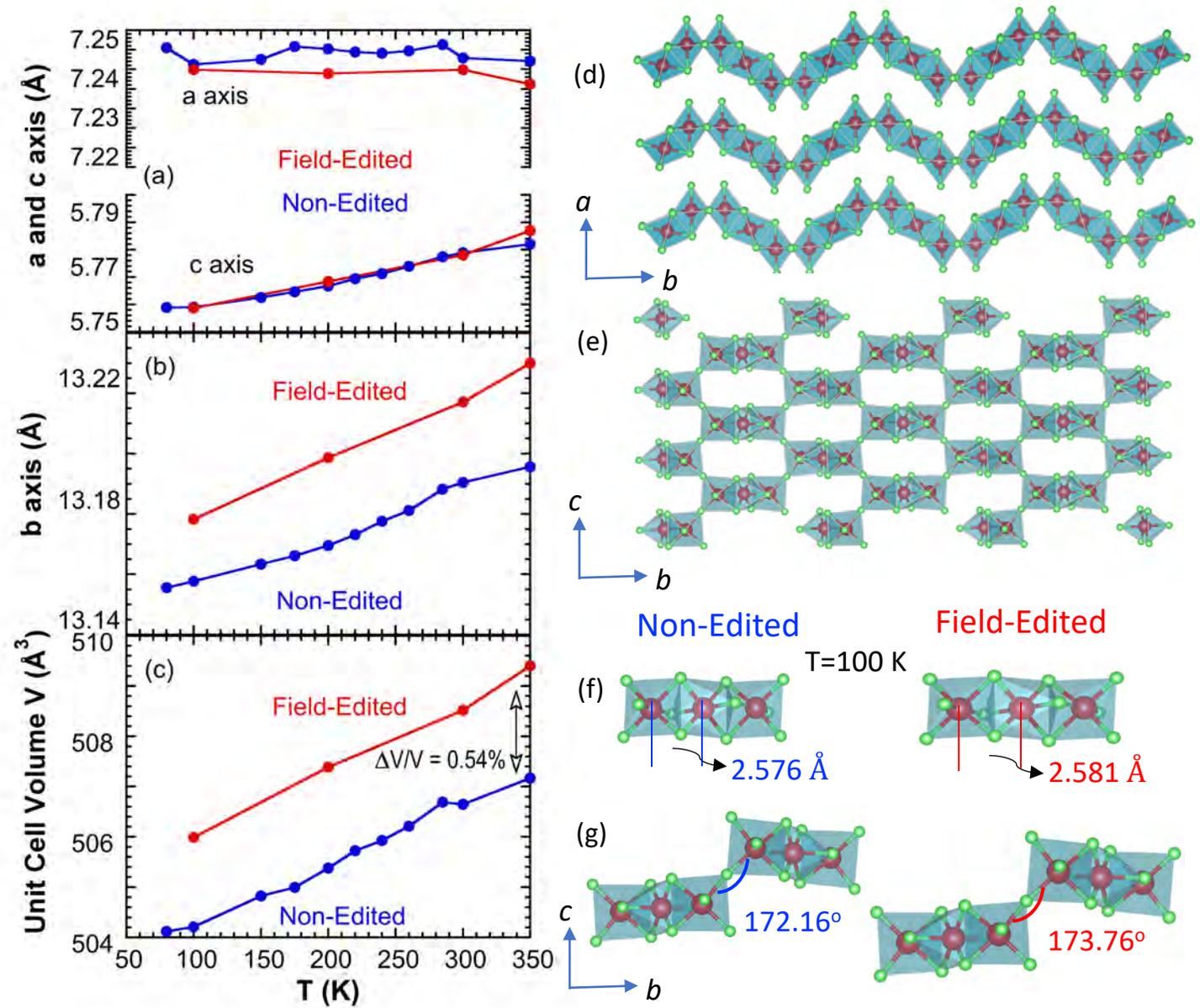

Fig.2

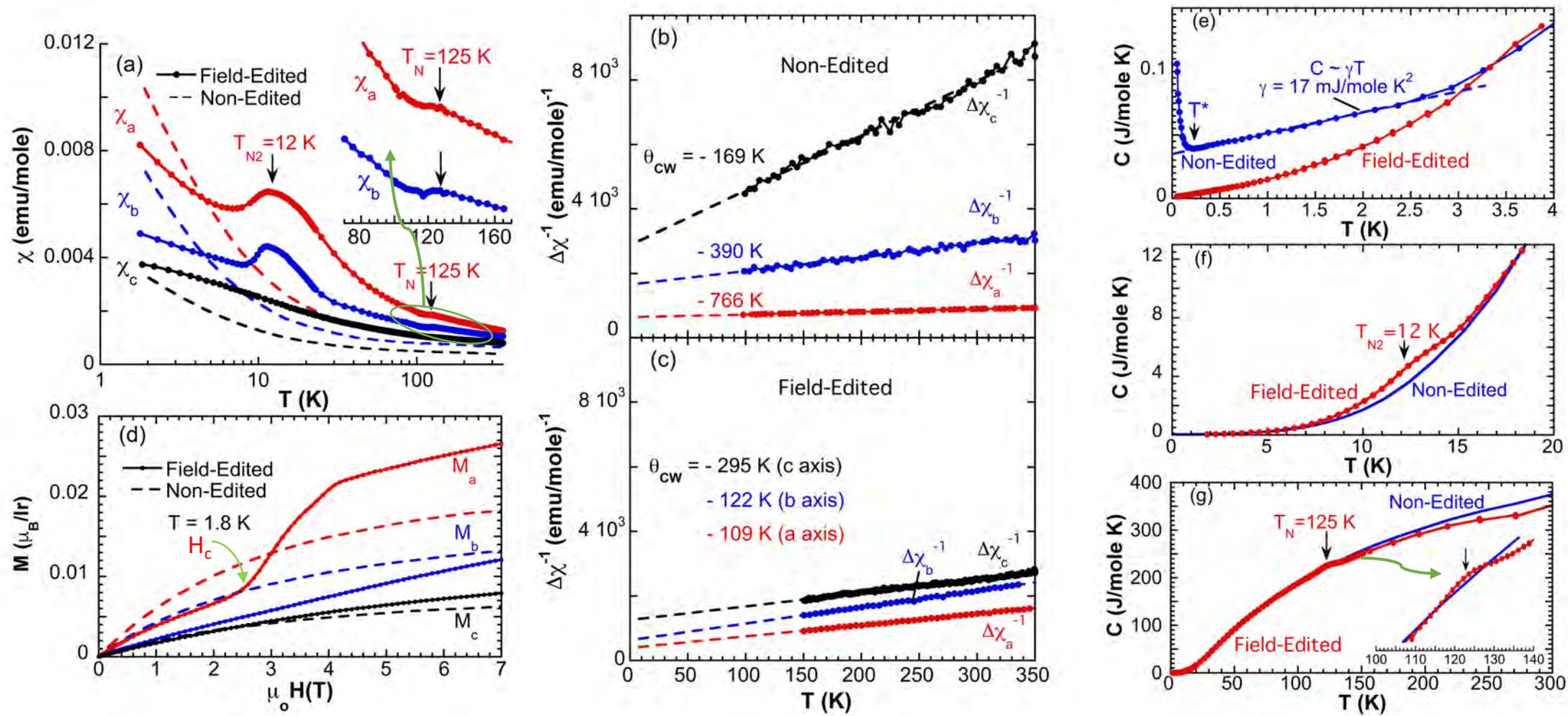

Fig. 3

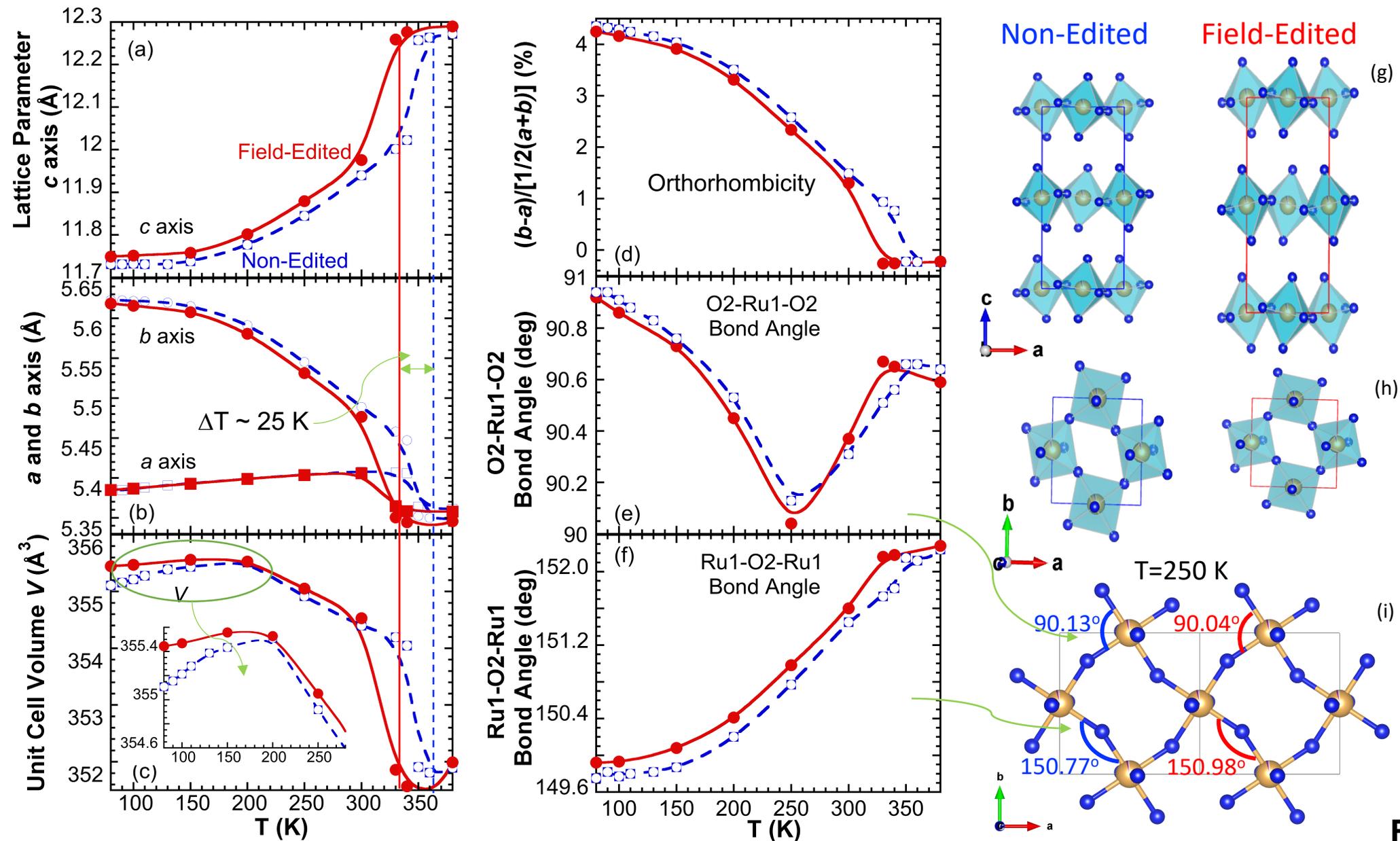

Fig. 4

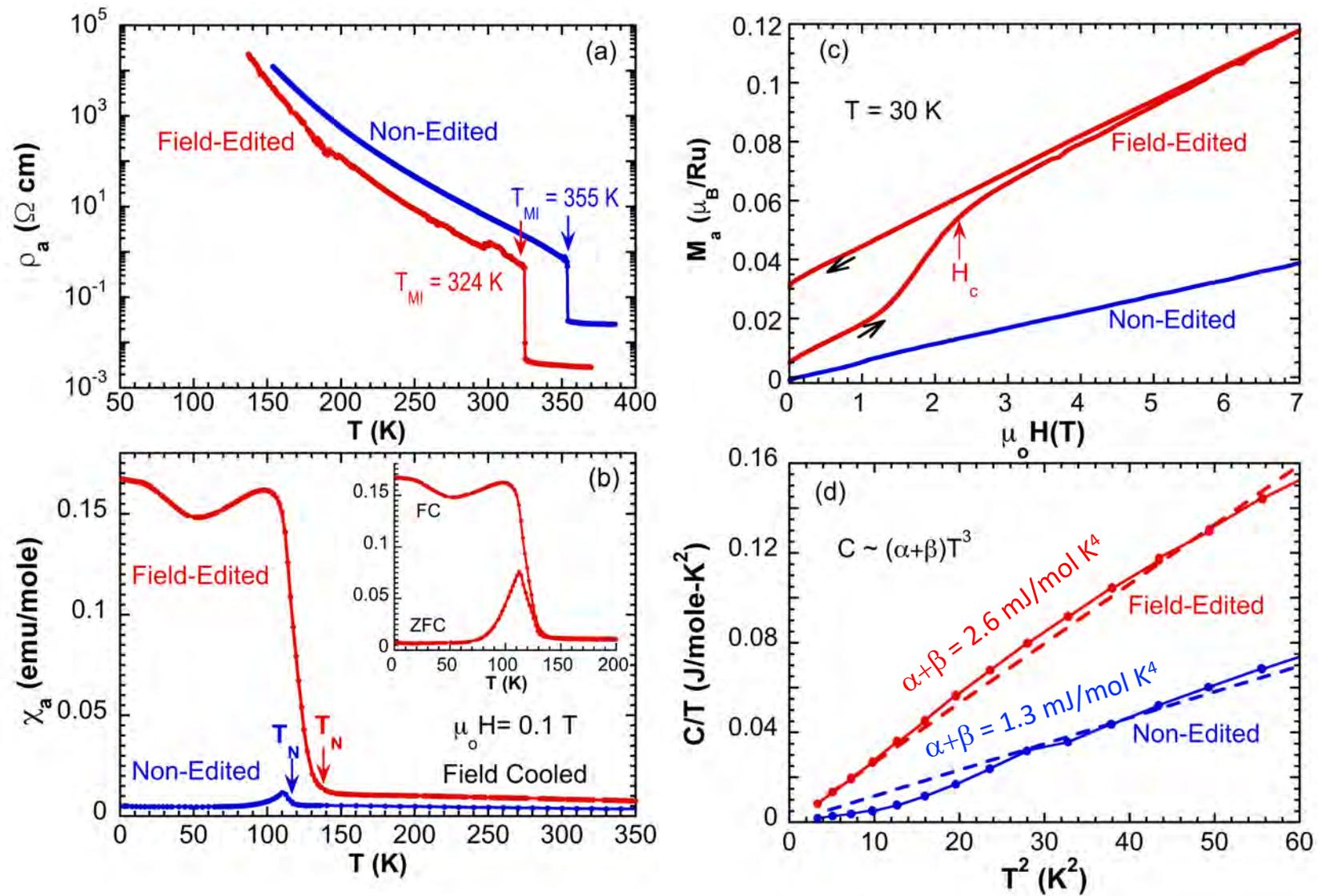

Fig. 5

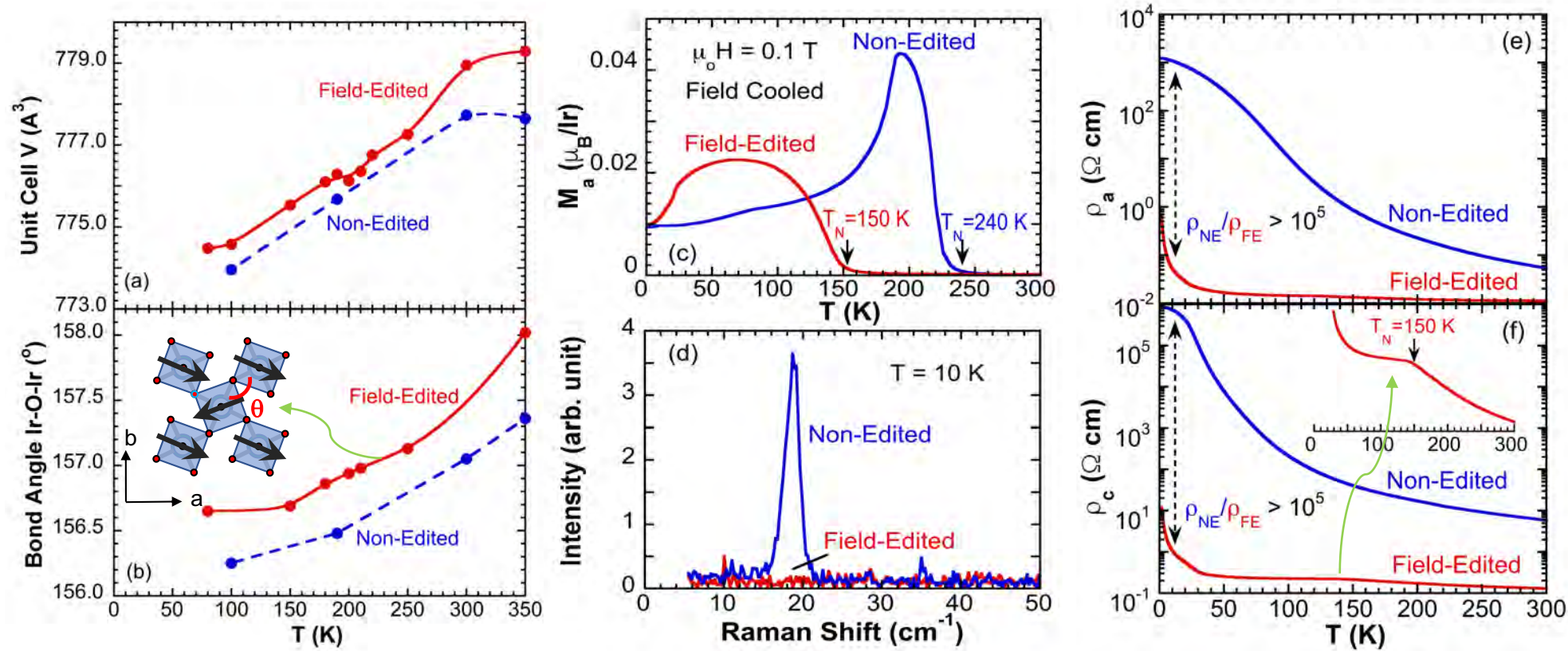

**Fig.6**